\documentclass{article}

\usepackage[utf8]{inputenc} 
\usepackage[T1]{fontenc}    
\usepackage{hyperref}       
\usepackage{url}            
\usepackage{booktabs}       
\usepackage{amsfonts}       
\usepackage{nicefrac}       
\usepackage{microtype}      
\usepackage{float}
\usepackage{natbib}
\usepackage{doi}
\usepackage{xcolor}
\usepackage{lineno}
\usepackage{amsmath}
\usepackage{enumitem}
\usepackage{changes}
\usepackage{algorithm}
\usepackage{url}
\newcommand\myshade{85}
\colorlet{mycitecolor}{violet}
\colorlet{myurlcolor}{blue}

\hypersetup{
  citecolor  = mycitecolor!\myshade!black,
  urlcolor   = myurlcolor!\myshade!black,
  colorlinks = true,
}

\usepackage{color}

\usepackage{times}
\usepackage{latexsym}
\usepackage{appendix}

\usepackage[noabbrev,capitalize]{cleveref}
\crefname{lstlisting}{listing}{listings}

\newcommand{\codefont}{\fontfamily{lmtt}\selectfont}
\usepackage{soul}
\definecolor{aigold}{RGB}{244,210, 1} 
\definecolor{aigreen}{RGB}{245, 255, 249}

\definecolor{humanpurple}{RGB}{235, 222, 240} 

\definecolor{commentgray}{RGB}{86, 101, 115}

\definecolor{aired}{RGB}{255,180,181}

\usepackage{listings}
\usepackage{parcolumns}
\lstdefinestyle{datalogstyle}{
	basicstyle={\codefont\small},  
	xleftmargin={6pt},
        xrightmargin={6pt},
        breakindent=0pt,
	frame=tb,
	stepnumber=1,
	firstnumber=1,
	numberfirstline=true,
	tabsize=2,
	showtabs=false,
	showspaces=false,
	showstringspaces=false,
	extendedchars=true,
	breaklines=true,
	columns=fullflexible,
	keepspaces=true,
	escapeinside={@}{@},
	firstnumber=last,
	captionpos=b,
	commentstyle=\color{black!65},
	numberstyle=\tiny\color{black!65},
	stringstyle=\color{codepurple},
	breakatwhitespace=false, 
	keepspaces=true,                 
	numbersep=5pt,                  
	showspaces=false,                
	showstringspaces=false,
	showtabs=false,
	aboveskip={0.8\baselineskip},
	belowskip={0.2\baselineskip},
	backgroundcolor=\color{aigreen},
}
\usepackage{tabularx, booktabs}
\newcolumntype{C}{>{\centering\arraybackslash}X}
\newcolumntype{R}{>{\raggedleft\arraybackslash}X}
\newcolumntype{S}{>{\raggedleft\arraybackslash\hsize=.5\hsize}X}

\usepackage{arydshln}
\usepackage{booktabs}
\usepackage{multirow}

\usepackage{pifont}  %
\usepackage{makecell}

\usepackage[T1]{fontenc}

\usepackage[utf8]{inputenc}

\usepackage{microtype}

\usepackage{fdsymbol}
\usepackage{booktabs}
\usepackage{graphicx}
\usepackage{makecell}
\usepackage{multirow}
\usepackage{longtable}
\usepackage{CJKutf8}
\usepackage{float}
\usepackage{algorithm}
\usepackage{algpseudocode}
\usepackage[frozencache, cachedir=minted-cache]{minted}
\usepackage[normalem]{ulem}

\usepackage{bm}
\renewcommand{\vec}[1]{{\boldsymbol{\mathbf{#1}}}}   %

\newcommand{\set}[1]{\mathcal{#1}}
\usepackage{tikz}
\usetikzlibrary{shapes.geometric}

\title{DB-GPT: Empowering Database Interactions with Private Large Language Models}

\date{} 					

\author{%
  Siqiao Xue$^{\diamondsuit}$, Caigao Jiang$^{\diamondsuit}$, Wenhui Shi$^{\diamondsuit}$, Fangyin Cheng$^{\varheartsuit}$, Keting Chen$^{\diamondsuit}$, Hongjun Yang$^{\diamondsuit}$, \\ 
  \textbf{Zhiping Zhang$^{\heartsuit}$, Jianshan He$^{\diamondsuit}$, Hongyang Zhang$^{\vardiamondsuit}$, Ganglin Wei$^{\diamondsuit}$, Wang Zhao, } \\
  \textbf{Fan Zhou$^{\diamondsuit}$, Danrui Qi$^{\clubsuit}$, Hong Yi, Shaodong Liu$^{\spadesuit}$, Faqiang Chen$^{\diamondsuit,*}$} \\
  $^{\diamondsuit}$Ant Group, $^{\heartsuit}$Alibaba Group, $^{\varheartsuit}$ JD Group, $^{\spadesuit}$Meituan\\
  $^{\vardiamondsuit}$Southwestern University of Finance and Economics, China\\ $^{\clubsuit}$Simon Fraser University, Canada\\
  \texttt{\{siqiao.xsq,caigao.jcg,faqiang.cfq\}@antgroup.com}\\
}

\usepackage[noabbrev,capitalize]{cleveref} %

\usepackage{pifont}  %
\usepackage{makecell}
\newcommand{\cmark}{\textcolor{green}{\ding{51}}}%
\newcommand{\xmark}{\textcolor{red}{\ding{55}}}%

\usepackage[preprint]{neurips_2023}

\begin{document}
\maketitle
\def\thefootnote{*}\footnotetext{Corresponding author.}
\def\thefootnote{\arabic{footnote}}
\begin{abstract}

The recent breakthroughs in large language models (LLMs) are positioned to transition many areas of software. Database technologies particularly have an important entanglement with LLMs as efficient and intuitive database interactions are paramount. In this paper, we present DB-GPT, a revolutionary and production-ready project that integrates LLMs with traditional database systems to enhance user experience and accessibility. DB-GPT is designed to understand natural language queries, provide context-aware responses, and generate complex SQL queries with high accuracy, making it an indispensable tool for users ranging from novice to expert. The core innovation in DB-GPT lies in its private LLM technology, which is fine-tuned on domain-specific corpora to maintain user privacy and ensure data security while offering the benefits of state-of-the-art LLMs. We detail the architecture of DB-GPT, which includes a novel retrieval augmented generation (RAG) knowledge system, an adaptive learning mechanism to continuously improve performance based on user feedback and a service-oriented multi-model framework (SMMF) with powerful data-driven agents. Our extensive experiments and user studies confirm that DB-GPT represents a paradigm shift in database interactions, offering a more natural, efficient, and secure way to engage with data repositories. The paper concludes with a discussion of the implications of DB-GPT framework on the future of human-database interaction and outlines potential avenues for further enhancements and applications in the field. The project code is available at {\small \url{https://github.com/eosphoros-ai/DB-GPT}}. Experience DB-GPT for yourself by installing it with the instructions {\small\url{https://github.com/eosphoros-ai/DB-GPT#install}} and view a concise 10-minute video at {\small\url{https://www.youtube.com/watch?v=KYs4nTDzEhk}}.
\end{abstract}


\section{Introduction}



Large language models (LLMs) such as ChatGPT~\citep{gpt3} and GPT-4~\citep{gpt4} have showcased their remarkable capabilities in engaging in human-like communication and understanding complex queries, bringing a trend of incorporating LLMs in various fields~\citep{anil2023palm,gunasekar2023textbooks}. These models have been further enhanced by external tools, enabling them to search for relevant online information~\citep{nakano2021webgpt,xue2023weaverbird}, utilize tools~\citep{schick2023toolformer}, and create more sophisticated applications~\citep{langchain,wang2023enhancing,chu2023leveraging}. In the realm of databases, while traditional systems often demand a high degree of technical acumen and familiarity with domain-specific structural query languages (SQLs) for data access and manipulation, LLMs pave the way for natural language interfaces, enabling users to express through natural language queries and leading to more natural and intuitive database interactions.

Nonetheless, how to empower the database operations with LLMs to build powerful end-user applications still remains an open question. One straightforward approach, employed by most of existing works~\citep{langchain,zhou2023llm,hu2023chatdb}, is to directly providing commonly used LLMs, such as GPT-4, with instructions on how to interact via few-short prompting or in-context learning~\citep{wei2022emergent}. The advantages of this approach is, it is unlikely to over-fit to train data and is easy to adapt to new data while the disadvantages are the performance can be sub-optimal compared to the fine-tuned alternatives with median-sized LLMs~\citep{sun2023sqlpalm}. Moreover, to further facilitate the intelligent interactions with database, many works~\citep{langchain,Liu_LlamaIndex_2022,autogpt} have incorporated the LLM-powered automated reasoning and decision process (a.k.a., agent) into the database applications. However, the knowledge agents are usually task-specific instead of task agnostic, limiting their use to a large scale. Meanwhile, though being important, the privacy-sensitive setup for LLM-centric database interactions have been under-investigated. The previous efforts~\citep{Martinez_Toro_PrivateGPT_2023,h2ogpt2023} are mostly general-purpose and not specially designed for database operations.

\begin{table*}
\scalebox{0.85}{
\begin{tabular}{ l c c c c c }
\toprule
& LangChain & LlmaIndex & PrivateGPT & ChatDB & DB-GPT\\
& \citep{langchain} & \citep{Liu_LlamaIndex_2022}  &     \citep{Martinez_Toro_PrivateGPT_2023}   &  \citep{hu2023chatdb}      &         \\
\midrule
Multi-LLM integration & \cmark & \cmark & \xmark & \cmark & \cmark \\
Text-to-SQL fine-tuned     & \xmark       & \cmark       & \xmark  & \xmark & \cmark                  \\
Multi-agent strategies  & \cmark  & \cmark    & \xmark     & \xmark    & \cmark     \\
Data privacy and security  & \cmark  & \xmark    & \cmark     & \xmark    & \cmark     \\
Multi-source knowledge  & \cmark  &  \cmark & \xmark & \xmark    & \cmark      \\
Bilingual queries  & \xmark  &  \xmark & \xmark & \cmark    & \cmark      \\
Generative data analytics  & \xmark  &  \xmark & \xmark & \xmark    & \cmark      \\
\bottomrule
\end{tabular}}
\caption{Comparative summary of competing approaches on various dimensions.}
\label{tab:comparison}
\vspace{-2mm}
\end{table*}

In this work, we introduce DB-GPT, an intelligent and production-ready project for LLM-augmented applications to ingest, structure, and access data with privatization technologies. DB-GPT harnesses not only the inherent natural language understanding and generation capabilities of LLMs but also continuously optimizes the data-driven engine through the agent and plugin mechanism. See \cref{tab:comparison} for a comparative summary of competitors. To summarize, DB-GPT has the following distinct merits:
\begin{figure}[ht]
\begin{center}
\centerline{\includegraphics[width=\columnwidth]{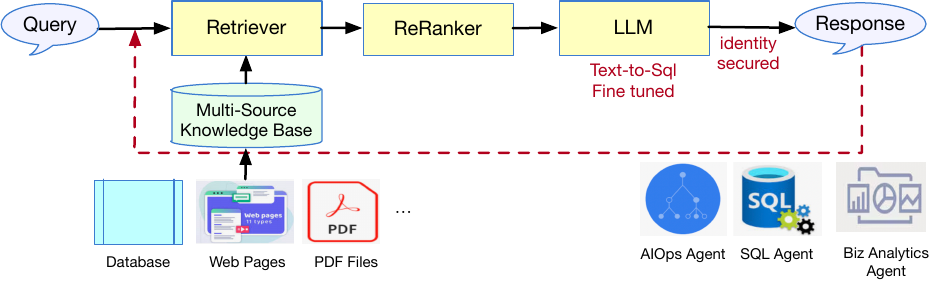}}
\caption{The architecture of DB-GPT}
\label{fig:dbgpt}
\end{center}
\vskip -0.2in
\end{figure}

\begin{itemize}[leftmargin=*]
    \item \textbf{Privacy and security protection.} DB-GPT allows users to deploy on personal devices or local servers and run even in scenarios without Internet connection. No data leaves the execution environment at any point, completely eliminating the risk of data leakage. In addition, proxy de-identification~\citep{wang2016identity} techniques are applied in data processing modules, which acts as an intermediary that obscures personal identifiers from datasets, thereby mitigating the risks of unauthorized access and exploitation of private information.

    
    \item \textbf{Multi-source knowledge base question \& answering optimized.} In contrast to classical works~\citep{lan2022complex} of knowledge base question \& answering (KBQA), DB-GPT builds a pipeline that ingests multi-source unstructured data (PDF's, web pages, images, etc) into intermediate representations, stores them in a structured knowledge base, retrieves the most relevant pieces,
    and generates a comprehensive natural language response given a query. The pipeline is efficiency-optimized, flexible in generation and accepts bilingual queries.
    

    \item \textbf{Text-to-SQL fine-tuned.} 
    To further enhance the generation capability, DB-GPT fine-tuned several commonly used LLMs (e.g., Llama-2~\citep{touvron2023llama}, GLM~\citep{zeng2022glm}) for Text-to-SQL tasks. DB-GPT significantly lowers the barriers to users without the expertise of SQL when interacting with data. To the best of our knowledge, among related works, only LlamaIndex~\citep{Liu_LlamaIndex_2022} integrates such fine-tuned alternatives but it is not optimized for bilingual queries.
    
    

    \item \textbf{Knowledge agents and plugins integrated.} An ``agent'' is an automated reasoning and decision engine. As a production-ready project, DB-GPT enables the development and application of conversational agents with advanced data analytics, where these automated decisions help interactive use cases over the data. It also offers a variety of plugins of query and retrieval services to use as tools for interaction with data.
    
    
\end{itemize}

We rigorously evaluate DB-GPT on various benchmark tasks, such as Text-to-SQL and KBQA. Furthermore, we conduct case studies and surveys to assess the usability and preferences. DB-GPT outperforms the competitors for most of the dimensions.

\section{SYSTEM DESIGN}
The overall pipeline of DB-GPT is depicted in \cref{fig:dbgpt}. While building upon the general Retrieval-Augmented Generation (RAG) framework~\citep{langchain,Liu_LlamaIndex_2022,xue2023weaverbird}, our DB-GPT system integrates our novel training and inference techniques, which
significantly enhance its overall performance and efficiency. In this
section, we delineate the design of each phase, including the model
architecture as well as training and inference paradigms.

\subsection{Multi-source RAG for QA}
\label{sec:rag}

\begin{figure}[ht]
\begin{center}
\centerline{\includegraphics[width=\columnwidth]{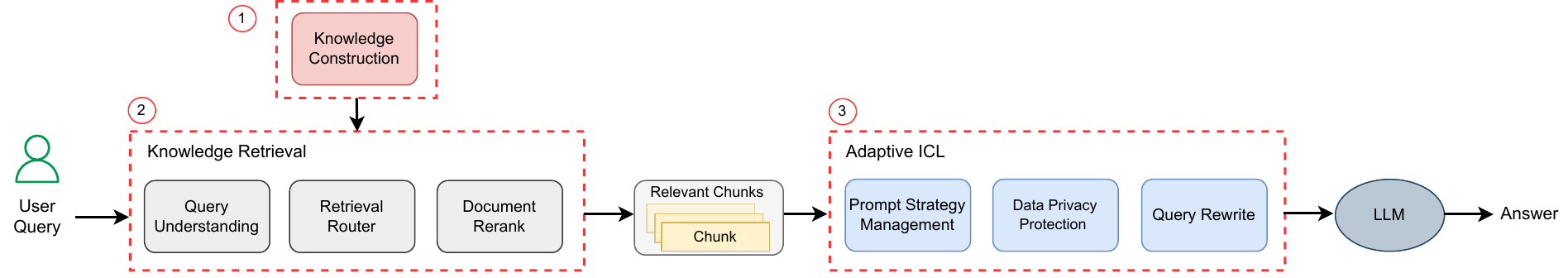}}
\caption{The detailed RAG architecture in DB-GPT}
\label{fig:rag_arch}
\end{center}
\vskip -0.2in
\end{figure}

While LLMs are usually trained on enormous bodies of open sourced or other parties' proprietary data, RAG~\citep{Lewis2020RetrievalAugmentedGF} is a technique for augmenting LLMs' knowledge with additional and often private data. Shown in \cref{fig:rag_arch}, our RAG pipeline consists of three stages: knowledge construction, knowledge retrieval and adaptive In-Contextual Learning (ICL)~\citep{Dong2022ASO} strategies.

\begin{figure}[ht]
\centering
\begin{minipage}[t]{0.325\columnwidth}
\centering
\includegraphics[width=\columnwidth,height=0.7\columnwidth]{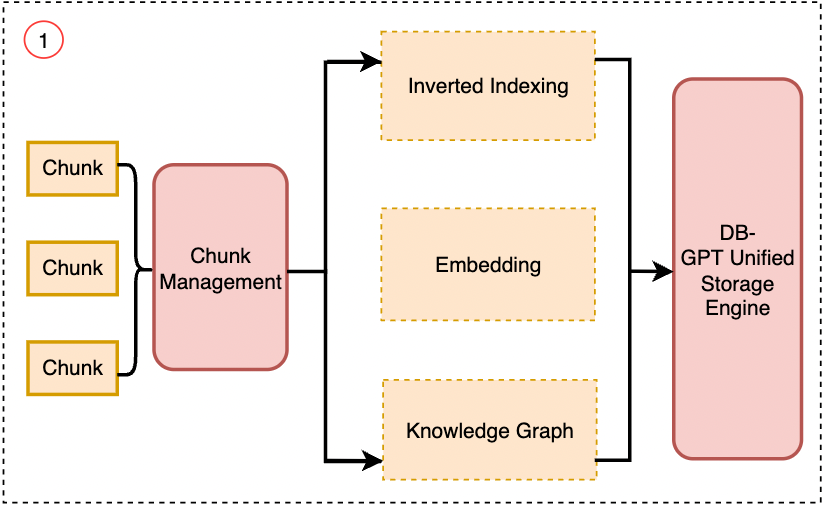}
\caption{The pipeline of knowledge construction}
\label{fig:kb_construct}
\end{minipage}
\begin{minipage}[t]{0.325\columnwidth}
\centering
\includegraphics[width=\columnwidth,height=0.7\columnwidth]{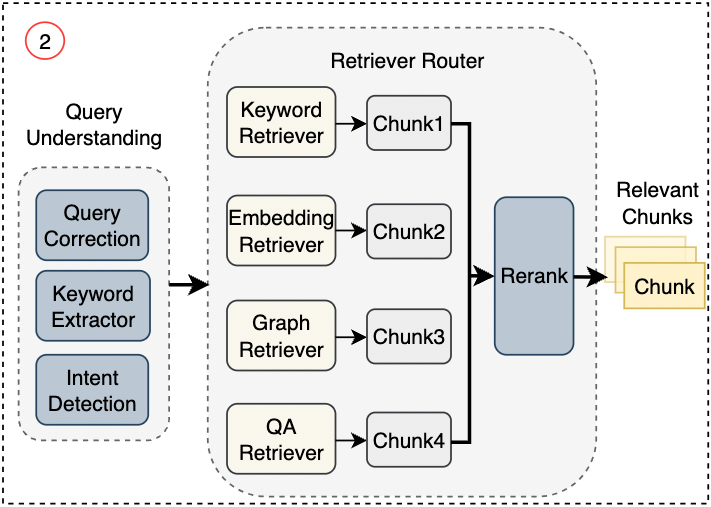}
\caption{The pipeline of knowledge retrieval}
\label{fig:retro}
\end{minipage}
\begin{minipage}[t]{0.325\columnwidth}
\centering
\includegraphics[width=\columnwidth,height=0.7\columnwidth]{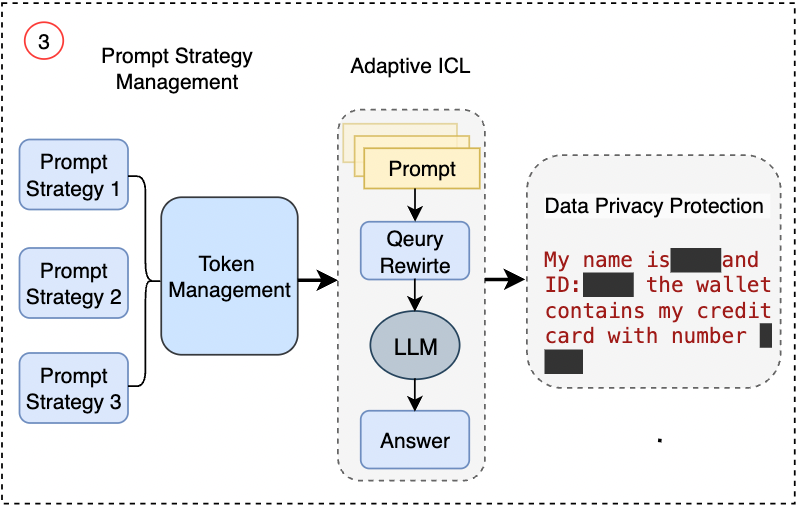}
\caption{The pipeline of adaptive ICL and response generation}
\label{fig:icl}
\end{minipage}
\end{figure}

\paragraph{Knowledge Construction.} Our knowledge base $\set{K}$ is a collection of documents from various sources $\vec{d}^{\text{loc}}_1, \ldots, \vec{d}^{\text{loc}}_N$ where the number of documents $N$ is large. Following \citet{langchain}, we split each document $\vec{d}_n$ into multiple paragraphs $\vec{p}^{\text{loc}}_{n,1},\ldots, \vec{p}^{\text{loc}}_{n,M_n}$, where $M_n$ (and $m$ below) denotes the index of paragraph for the $n$-th document, and embed each paragraph into a multidimensional embedding $\vec{e}^{\text{loc}}_{n,m}$ through a neural encoder $f_{\text{key}}$. It is worth noting that, in addition to the existing vector-based knowledge representation, shown in \cref{fig:kb_construct}, DB-GPT also incorporates inverted index and graph index techniques to make it easy to accurately find contextually relevant data.

\paragraph{Knowledge Retrieval.} Illustrated in \cref{fig:retro}, when a language query $\vec{x}$ comes, it is embedded into a vector $\vec{q}$ through another neural encoder $f_{\text{query}}$ and we retrieve the top $K$ relevant paragraphs from the knowledge base, where $K$ is a hyperparameter. DG-GPT supports various retriever models, e.g., EmbeddingRetriever, which retrieve according to their cosine similarities, i.e., $\vec{q}^{\top}\vec{e}/\|\vec{q}\|\|\vec{e}\|$, KeywordRetriever, which match keywords instead of whole sentences. In the following paragraphs, we assume EmbeddingRetriever is used by default.

\paragraph{Learning to Embed and Search.} Following~\citet{xue2023weaverbird}, we confidently consider a higher similarity to signify a more relevant paragraph due to the
training of the encoders $f_{\text{key}}$ and $f_{\text{query}}$. Their optimization is clarified in \cref{sec:encoder}. 
Intuitively, we want the dot products $\vec{q}^{\top} \vec{e}$ to be relatively large for the query-paragraph pairs that are actually relevant. Our encoders use Multilingual-E5-base model architecture~\citep{wang2022text} as we support billingual encoding documents. 

\paragraph{Adaptive ICL and Generation by LLM.} In this phase, our system performs the ICL~\citep{Dong2022ASO} for response generation: it ranks the $K$ search results based on their cosine similarities with the query, then plugs the top $J$ (where $J \le K$) results into the context part of the predefined prompt template and finally LLM generates a response. ICL is a technique used to improve LLMs' performance in handling contextual information by incorporating additional context during the training or inference phase. The whole process is shown in \cref{fig:icl}. ICL empowers language models with enhanced understanding of context, improved reasoning and inference skills, and tailored problem-solving capabilities. As the performance of ICL is sensitive to specific settings, including the prompting template, the
selection of in-context examples, and order of examples, and so on~\citep{Zhao2021Calibrate}, in our DB-GPT system, we offers several strategies to formulate prompting template (see \Cref{lst:en_template} for one example). 
 In addition, we apply the privacy protection measure to mask the personal information.



\begin{lstlisting}[caption={Prompt templates for LLM.},label={lst:en_template}]
Context information:
{CONTEXT_RETRO_1} 
@
\vdots
@
{CONTEXT_RETRO_K} 

 Based on the given information, please provide a concise and professional response to the user's question. If there are multiple questions in a query, please answer all of them. If the user's question includes keywords like 'recent' or 'latest' to indicate a recent time frame, pay attention to the correspondence between the current date and the date of the information. If a clear answer cannot be determined, respond with "Unable to answer the question based on the information provided". You MUST respond in the same language as the question!

The question is: {QUESTION}.
\end{lstlisting}
\vspace{-0.3cm}

\subsection{Deploy and Inference: Service-oriented Multi-model Framework}
\label{sec:smmf}

Model-as-a-Service (MaaS) is a cloud-based AI approach that provides developers and businesses with access to pre-built, pre-trained machine learning models. In DB-GPT,  in order to streamline model adaptation, enhance the efficiency, and optimize the performance of model deployment, we present the Service-oriented Multi-model Framework (SMMF), which provides a fast and easy-to-use platform for the deployment and inference for Multi-LLMs.

SMMF consists of two principal components, namely the model inference layer and the model deployment layer. Specifically, the model inference layer is designed for accommodating various LLM inference platforms, including vLLM~\citep{kwon2023efficient}, HuggingFace Transformers (HF)~\citep{hf}, Text Generation Inference (TGI)~\citep{tgi}, and TensorRT~\citep{tensorrt}. The model deployment layer serves as an intermediary between the underlying inference layer and the upper-level model serving functionalities. 

\paragraph{Deployment Layer.}
Within the context of the model deployment framework layer, a suite of integral elements can be identified. A duo composed of the API server alongside the model handler is tasked with providing potent model serving functions to the application stratum. Occupying a central position, the model controller is entrusted with the governance of metadata while also operating as the nexus for the extensive deployment architecture. Additionally, the model worker is of paramount importance, establishing a direct connection with the inference apparatus and the foundational setting, thereby ensuring a proficient performance of the implemented models.



\subsection{Multi-agent Strategies}

DB-GPT supports several roles to interact with data, such as data analyst, software engineer and database architect, providing the entire process of database operations along with carefully orchestrated Standard Operating Procedures (SOPs). Inspired by MetaGPT~\citep{hong2023metagpt}, DB-GPT assigns distinct roles to individual agents, leveraging their strengths and specialties to tackle challenging tasks. It orchestrates collaboration between different LLM agents through a coordination mechanism, enabling them to communicate, share information, and collectively reason. Based on the Text-to-SQL fine-tuned LLM, DB-GPT enables the development and application of agents with advanced interaction ability with database. Besides, different from LlamaIndex, whose components offer more explicit, constrained behavior for specific use cases, DB-GPT empowers agents with stronger capability of general reasoning with less constraint. 

\subsection{DB Plugins}
\label{sec:db_plugings}

LLMs are undoubtedly powerful, yet they may not excel at every task. Instead of answering the questions directly, an LLM can perform multiple steps to gather the relevant information by incorporating plugins (also known as tools)\footnote{In this paper, we interchangeably use these two terms.}. Different from general-purpose plugins~\citep{schick2023toolformer}, DB-GPT's plugins are predominantly rooted in database interaction modes. This design facilitates querying databases through natural language, streamlining user query expressions while reinforcing LLMs' query comprehension and execution abilities. The database interaction mode comprises two components: the schema analyzer, which deciphers the schema into a structured representation comprehensible by LLMs, and the query executor, which executes SQL queries on the database based on LLMs' natural language responses. Besides, DB-GPT also integrates with third party services, such as web search proposed in WebGPT~\citep{nakano2021webgpt}, executes tasks on another platform without leaving the chat. Empowered with these plugins, DB-GPT is able to conduct several end-to-end data analysis problems with strong generative ability (we call it \emph{generative data analytics} in our context). See for illustrative examples.

\section{Models and Training}
\label{sec:model}

\subsection{Text-to-SQL Fine-Tuning}
\label{sec:text2sql}

Although LLMs,.e.g., CodeX~\citep{chen2021evaluating} and ChatGPT~\citep{liu2023comprehensive}, have shown successful results with ICL for Text-to-SQL, they still have a gap with the fine-tuned alternatives with median-sized LLMs~\citep{sun2023sqlpalm}. Therefore, it is necessary to adapt LLMs to domain specific Text-to-SQL data, so that LLMs can better
understand the format of prompt and yield further improved results.


\paragraph{Model Architecture.} We started with a pre-trained Qwen~\citep{qwen} that has been pre-trained using extensive English and Chinese corpora.

\paragraph{Dataset and Training.}
In our DB-GPT, we have designed a special module DB-GPT-Hub that encapsulates the pipeline of preprocessing records (via the tools introduced in \cref{sec:db_plugings}), model loading and fine-tuning. We fine-tune Qwen on Spider~\citep{sprder2018} train split with inputs including description of database and natural question (see \Cref{lst:prompt_finetune}), and the output is the target SQL. 


\begin{lstlisting}[caption={Format of the input for Text-to-SQL fine-tuning.},label={lst:prompt_finetune}]
{"instruction": "concert_singer contains tables such as stadium, singer, concert, singer_in_concert. Table stadium has columns such as stadium_id, location, name, capacity, highest, lowest, average. stadium_id is the primary key. Table singer has columns such as singer_id, name, country, song_name, song_release_year, age, is_male. singer_id is the primary key. Table concert has columns such as concert_id, concert_name, theme, stadium_id, year. concert_id is the primary key. Table singer_in_concert has columns such as concert_id, singer_id. concert_id is the primary key. The year of concert is the foreign key of location of stadium. The stadium_id of singer_in_concert is the foreign key of name of singer. The singer_id of singer_in_concert is the foreign key of concert_name of concert.", 

"input": "How many singers do we have?", 

"response": "select count(*) from singer"}
\end{lstlisting}


The full details of architecture and evaluation of fine-tuning LLMs for Text-to-SQL can be found in another paper that we will publish soon.

\subsection{Encoder in RAG}
\label{sec:encoder}
\paragraph{Model Architecture.} The key and query encoders $f_{\text{key}}$ and $f_{\text{query}}$ are initialized as Multilingual-E5-base (ME5) model architecture~\citep{wang2022text} as we support bilingual applications. 

Their optimization involves maximizing a well-defined objective:
\begin{align}
    \ell 
    = 
    \vec{q}^{\top} \vec{e}_{0}
    - \log \sum_{i=0}^{I} \exp\left( \vec{q}^{\top} \vec{e}_{i} \right),
    \label{eqn:encoder_optim}
\end{align}
where $\vec{e}_{0}$ is the embedding of the paragraph known to contain relevant information for the query, and the other $I$ embeddings $\vec{e}_{1}, \ldots, \vec{e}_{I}$ belong to a set of negative paragraphs (see \cref{sec:encoder} for how they are selected). 
By optimizing \cref{eqn:encoder_optim}, the dot products $\vec{q}^{\top} \vec{e}$ become relatively large for the query-paragraph pairs that are actually relevant. 

\paragraph{Dataset and Training.}  Following the work of \citet{xue2023weaverbird}, we use query-paragraph pairs to train the key and query encoders $f_{\text{key}}$ and $f_{\text{query}}$. The pairs are collected from DatabaseQA (see \cref{sec:rag_eval}): we sample $1,000$ query-response pairs as the positive pairs and for each of them, we randomly sample five negative responses from the entire pool of paragraphs. Finally, we collect $1000$ query-response pairs for training and evaluation. The chosen pairs were then divided into sets of $700$ training pairs, $100$ development pairs, and $200$ test pairs. 

We pass query-response pairs to the
model to yield a scalar score for each of the pair
and maximize the scores for the positive pairs while minimizing the scores for the negative pairs with the cross entropy loss.

\subsection{Implementation and Deployment Details.}
\label{sec:implementation}

\paragraph{Knowledge Base and WebUI.} For the implementation of RAG, we use the code from the  public  GitHub  repository at {\small \url{https://github.com/langchain-ai/langchain}}~\citep{langchain} with MIT License as the reference. For the implementation of WebUI, we develop by our own and release at the GitHub  repository at {\small \url{https://github.com/eosphoros-ai/DB-GPT-Web}} with MIT License.


\paragraph{Deployment Details.}
For demo and testing purposes, unless otherwise specified, our system is deployed on a server on Alibaba Cloud with 30G RAM, a 8 logical cores (Intel Xeon(Ice Lake) Platinum 8369B), and a NVIDIA A100 80G Tensor Core GPU.

\section{Experiments}


We present the experiments designed to evaluate the performance of the DB-GPT system, including the generation quality of Text-to-SQL responses (\cref{sec:text2sql}) and the QA performance of our proposed RAG mechanism (\cref{sec:rag}) and the efficiency performance of SMMF~\citep{sec:smmf}. We also provide qualitative results of generative data analytics (\cref{sec:db_plugings}).

\subsection{Text-to-SQL Evaluation}
\label{sec:exp_text2sql}

\paragraph{Dataset.} We evaluate Text-to-SQL methods on Spider~\citep{sprder2018} dataset. Spider is a large-scale cross-domain Text-to-SQL dataset, which contains 8659
instances in training split and 1034 instances in development split over 200 databases. Each instance consists of a natural language question on a specific database and its corresponding SQL query. In this paper, we use the development split Spider-dev for the purpose of evaluation as the test split is not released. Each instance is divided into different categories(easy, medium, hard, and extra) according to the complexity of the question. See \cref{app:text2sql_details} for details.

\paragraph{Metrics.} We follow the prior study~\citep{liu2023comprehensive}
to use execution accuracy (EX) as the metric. EX compares the execution output of the predicted SQL query with that
of the ground truth SQL query on some database instances. The higher EX is considered the better.

\textbf{Base LLMs.} We compare Qwen that we chosen for DB-GPT with Baichuan~\citep{baichuan2023baichuan2} that also well support bilingual texts.

\begin{table*}[tbh]
	\begin{center}
		\begin{small}
			\begin{sc}
				\begin{tabularx}{1.00\textwidth}{l *{1}{S}*{4}{R}}
					\toprule
					Model & \multicolumn{4}{c}{Metrics (EX)}  \\
					\cmidrule(lr){2-6}
					& Easy & Medium & Hard & Extra & Overall  \\
					\midrule
Qwen-7b-chat	& 0.395      &          0.256      &          0.138             &   0.042      &          0.235 \\		
 Qwen-7b-chat-sft &  \textbf{0.911} & \textbf{0.675} & \textbf{0.575} & \textbf{0.343}  & \textbf{0.662}    \\
 \midrule
Qwen-14b-chat & 0.871          &      0.632    &            0.368     &           0.181             &   0.573 \\		

 Qwen-14b-chat-sft &  \textbf{0.919}  & \textbf{0.744}   & \textbf{0.598} & \textbf{0.367}  & \textbf{0.701}    \\
  \midrule
 Baichuan2-7b-chat & 0.577  &              0.352        &        0.201               & 0.066      &          0.335   \\
 Baichuan2-7b-chat-sft & \textbf{0.891}  &   \textbf{0.637}   & \textbf{0.489}  & \textbf{0.331}       & \textbf{0.624}    \\
  \midrule
 Baichuan2-13b-chat &  0.581      &          0.413   &             0.264              &  0.187      &          0.392   \\

 Baichuan2-13b-chat-sft & \textbf{0.895}  &  \textbf{0.675}   & \textbf{0.580}  & \textbf{0.343}  &   \textbf{0.659}    \\
  
					\bottomrule
				\end{tabularx}
			\end{sc}
		\end{small}
	\end{center}
	\caption{Evaluation on Spider-dev datasets.}
	\label{tab:main_results_text2sql}
\end{table*}

\paragraph{Main Result.} \cref{tab:main_results_text2sql} shows the effectiveness of the Text-to-SQL fine-tuning pipeline of our DB-GPT system: for both Qwen and Baichuan, the fine-tuned version shows significant improvement compared to the original LLM measured by EX.






\subsection{RAG Evaluation}
\label{sec:rag_eval}
Following~\citep{Lewis2020RetrievalAugmentedGF}, we experiment with RAG in a wide range of open-domain QA tasks.

\paragraph{Dataset.} We construct two QA datasets: DatabaseQA and FinancialQA. For DatabaseQA we collect $1000$ public tutorials in PDFs from three representative database systems: OceanBase~\citep{ob},  MySQL~\citep{mysql} and MongoDB~\citep{mongodb}. For FinancialQA, we sample $1000$ documents published from research institutes. For each dataset, we construct ~$100$ questions for testing, where questions are annotated with difficulties by experts. See \cref{app:rag_details} for more details on dataset.

\paragraph{Metrics.} We have three experts rate each response with rating from $0-5$, where the higher score is consider the better answer, and take the average of them as the final score.

\paragraph{Base LLMs.} We use four LLMs: Qwen, Baichuan and two commercial LLMs: ChatGLM-Turbo~\citep{zeng2022glm} and ChatGPT3.5~\citep{gpt3} as the base model, respectively. For ChatGLM-Turbo and ChatGPT3.5, we directly call the APIs to run our task.

\paragraph{Main Results.} Illustrated in \cref{tab:main_results_rag_database} and \cref{tab:main_results_rag_fin}, there is no consistent winner across the datasets: ChatGPT-3.5 is the winner on DatabaseQA dataset while ChatGLM2-7b achieves the best performance on FinancialQA dataset. As DB-GPT integrates most of the popular open source and commercial LLMs, the users are able to choose the most suitable one for their own RAG tasks.

\begin{table*}[tbh]
	\begin{center}
		\begin{small}
			\begin{sc}
				\begin{tabularx}{1.00\textwidth}{l *{1}{S}*{4}{R}}
					\toprule
					Model & \multicolumn{4}{c}{Metrics (Average Score)}  \\
					\cmidrule(lr){2-5}
					& Easy & Medium & Hard & Overall  \\
					\midrule
Qwen-7b-chat	& 0.487 & 0.488 & 0.485 & 0.487 \\		
Baichuan2-7b-chat &  0.470 & 0.468 & 0.466 & 0.468    \\
ChatGLM-Turbo &  0.460 & 0.459 & 0.464 & 0.461   \\
ChatGPT-3.5 &  \textbf{0.663} & \textbf{0.644} & \textbf{0.628} & \textbf{0.645}   \\
					\bottomrule
				\end{tabularx}
			\end{sc}
		\end{small}
	\end{center}
	\caption{RAG Evaluation on DatabaseQA dataset.}
	\label{tab:main_results_rag_database}
\end{table*}

\begin{table*}[tbh]
	\begin{center}
		\begin{small}
			\begin{sc}
				\begin{tabularx}{1.00\textwidth}{l *{1}{S}*{4}{R}}
					\toprule
					Model & \multicolumn{4}{c}{Metrics (Average Score)}  \\
					\cmidrule(lr){2-5}
					& Easy & Medium & Hard & Overall  \\
					\midrule
Qwen-7b-chat	& 0.829 & 0.824 & 0.819 & 0.824 \\		
Baichuan2-7b-chat &  0.897 & 0.893 & 0.895 & 0.895    \\
ChatGLM-Turbo &  \textbf{0.910} & \textbf{0.905} & \textbf{0.900} & \textbf{0.905}    \\
ChatGPT-3.5 &  0.903 & 0.899 & 0.898 & 0.900    \\
					\bottomrule
				\end{tabularx}
			\end{sc}
		\end{small}
	\end{center}
	\caption{RAG Evaluation on FinancialQA dataset.}
	\label{tab:main_results_rag_fin}
\end{table*}

\subsection{SMMF Evaluation}
\label{sec:exp_smmf}
Clarified in \cref{sec:smmf}, our DB-GPT integrates vLLM as the main inference framework.

\paragraph{Dataset.} The test is performed on a server with
629G RAM, a hard drive with 1TB HDD, a 40 logical cores (Intel Xeon Processor (Skylake, IBRS)) CPU at 2992.953MHz, and a NVIDIA A100-PCIE GPU with 40G GPU RAM. Across all the experiments, we use the same prompt with $8$ tokens as the input while setting the output length to be $256$ tokens. 

\paragraph{Metrics.} We use the following three metrics: 
\begin{itemize}[leftmargin=*]
    \item First Token Latency (FTL): measured in milliseconds, it represents the time spent from the moment the DB-GPT model deployment framework receives a request to the point when the inference framework decodes the first token. 
    \item Inference Latency (IL): measured in seconds, it represents the time spent from the moment the DB-GPT model deployment framework receives a request to the point when the inference framework decodes the complete response. 
    \item Throughput: the total number of tokens processed by the DB-GPT model deployment framework per second, across all requests.
\end{itemize}

\paragraph{Base LLMs.} Same as in \cref{sec:text2sql}, we use Qwen and Baichuan as the base LLMs for the experiment.

\paragraph{Main Results.} Seen in \cref{tab:main_results_smmf_qwen} and \cref{tab:main_results_smmf_baichuan}, the results indicate that the use of the vLLM framework for model inference significantly increases the throughput of the model, while substantially reducing both the first token latency and the overall inference latency. Moreover, as the number of concurrent users rises, the performance improvements gained from utilizing the vLLM framework for inference become particularly pronounced. As a result, DB-GPT choose to integrate vLLM as the default inference framework used for SMMF.

\begin{table*}[tbh]
	\begin{center}
		\begin{small}
			\begin{sc}
				\begin{tabularx}{1.00\textwidth}{l *{1}{S}*{1}{S}*{3}{R}}
					\toprule
					Model & \# Ccr & Platform & \multicolumn{3}{c}{Metrics}  \\
					\cmidrule(lr){4-6}
					& & & FTL(ms) & IL(s) & Throughput (tokens)  \\
					\midrule		
 Qwen-7b-Chat & 4 & vLLM &  \textbf{22.5}&	\textbf{4.0}&	\textbf{258.9}   \\
 Qwen-7b-Chat & 4 & HF &  765.7 & 97.6 & 10.7   \\
 \midrule
 Qwen-7b-Chat & 16 & vLLM &  \textbf{23.1} &	\textbf{4.1}	& \textbf{258.7}   \\
 Qwen-7b-Chat & 16 & HF &  1152.0 & 138.9 & 9.2   \\
 \midrule
 Qwen-7b-Chat & 32 & vLLM &  \textbf{23.3}&	\textbf{4.2}&	\textbf{289.2}   \\
 Qwen-7b-Chat & 32 & HF &  1059.2 & 127.1 & 10.1   \\      					\bottomrule
				\end{tabularx}
			\end{sc}
		\end{small}
	\end{center}
	\caption{Evaluation on SMMF with Qwen as the base LLM.}
	\label{tab:main_results_smmf_qwen}
\end{table*}

\begin{table*}[tbh]
	\begin{center}
		\begin{small}
			\begin{sc}
				\begin{tabularx}{1.00\textwidth}{l *{1}{S}*{1}{S}*{3}{R}}
					\toprule
					Model & \# Ccr & Platform & \multicolumn{3}{c}{Metrics}  \\
					\cmidrule(lr){4-6}
					& & & FTL (ms) & IL (s) & Throughput (tokens) \\
					\midrule		
 Baichuan-7b-Chat & 4 & vLLM &  \textbf{54.7}&	\textbf{5.2}&	\textbf{201.7}   \\
 Baichuan-7b-Chat & 4 & HF &  688.5 & 70.8 & 14.7   \\
 \midrule
 Baichuan-7b-Chat & 16 & vLLM &  \textbf{156.2} &	\textbf{7.1}	& \textbf{588.2}   \\
 Baichuan-7b-Chat & 16 & HF &  2911.7 & 985.4 & 4.2   \\
 \midrule
 Baichuan-7b-Chat & 32 & vLLM &  \textbf{380.0}&	\textbf{9.6}&	\textbf{870.2}   \\
 Baichuan-7b-Chat & 32 & HF &  6786.6 & 1630.7 & 5.1   \\      					\bottomrule
				\end{tabularx}
			\end{sc}
		\end{small}
	\end{center}
	\caption{Evaluation on SMMF with Baichuan as the base LLM.}
	\label{tab:main_results_smmf_baichuan}
\end{table*}

\section{Related Work}

\textbf{LLM for Databases.} The emergence of LLMs has revolutionized various application domains. In recent years, researchers have explored the potential of LLMs in the context of databases, aiming to revolutionize the way the users interact with and query databases. The most relevant works are LangChain~\citep{langchain} and LllmaIndex~\citep{Liu_LlamaIndex_2022}. Our DB-GPT differs from them mainly in terms of bilingual queries and generative data analytics integration. Among other relevant works, PrivateGPT~\citep{Martinez_Toro_PrivateGPT_2023} focuses on security and privacy setup of the LLM-based database applications while ChatDB~\citep{hu2023chatdb} mainly addresses the LLM-based SQL generation and reasoning problem with a symbolic memory framework. A comparison of
our DB-GPT system to other competitive approaches is summarized in \cref{tab:comparison}.

\textbf{LLM Agent.} An agent takes in a user input or query and can make internal decisions for executing that query in order to return the correct result. Recent developments in LLMs~\citep{yao2023react} and LLM tooling~\citep{autogpt,hong2023metagpt} have popularized the concept of agents. Compared to agent frameworks used in Langchain and LlamaIndex, DB-GPT has implemented agents that are with less constraints and more task agnostic, supported by the strong reasoning ability of the fine-tuned model.

\textbf{Knowledge Base Question and Answer and Retrieval-Augmented Generation.} Knowledge base question answering (KBQA) plays a crucial role in leveraging the substantial knowledge stored in knowledge bases and making it accessible to users~\citep{lan2022complex,cao-etal-2022-kqa}. LLMs' remarkable generalization with minimal demonstrations~\citep{shi2023language} hints at its potential in KBQA. Our DB-GPT system aligns with works on LLM-based database applications, which involves enhancing language models by incorporating external datastores, such as PDF's, web pages, Google Docs, etc. In line with LllmaIndex, our DB-GPT system implements a robust indexing structure, categorising documents into nodes for efficient information retrieval. The language generation process of combining retrieved external knowledge and pre-trained parametric knowledge is called retrieval-augmented generation (RAG)~\citep{Lewis2020RetrievalAugmentedGF}, which has been widely used in knowledge-intensive tasks and database applications. In addition to the standard pipeline of RAG, DB-GPT provides various types of bilingual text splitting, embedding, ranking methods, which is more flexible than competitors. 

\textbf{Deployment Platform for LLM-based Applications.} The deployment platforms of LLM-based application can categorized into two primary groups: distributed systems and cloud platforms. Distributed systems distribute LLMs across multiple nodes to enhance performance and reliability through network coordination and load balancing. Conversely, cloud platforms host LLMs on cloud servers, offering simplicity and ease of management through user-friendly interfaces or APIs. FastChat~\citep{zheng2023judging} stands out in this context, enabling the deployment of LLMs on any cloud platform. It provides a Web UI for streamlined model and task management and supports OpenAI-compatible RESTful APIs for seamless integration. Additionally, SkyPilot~\citep{skypilot} offers a diverse range of billing strategies and optimization techniques to enhance cost-efficiency and GPU utilization. Cloud platforms have become the preferred and flexible choice for LLM deployment, eliminating concerns about underlying architecture and details. While supporting both types of platforms, DB-GPT allows users to deploy on personal devices or local servers and run even in scenarios without Internet connection, which ensures data security and privacy.

\textbf{Text-to-SQL Fine-Tuning.}
Text-to-SQL aims to automate the process of generating SQL queries for databases from natural language text. It is a long-standing challenge, crucial to enhance database accessibility without requiring
expertise of SQL. LLMs, such as GPT-4~\citep{gpt4}, PALM~\citep{chowdhery2022palm} and Llama-2~\citep{touvron2023llama}, have shown significant achievements with few-shot prompting or in-context learning on this task and the performance can be further improved by fine-tuning~\citep{sun2023sqlpalm}. Compared to several competitors, e.g., LangChain, PrivateGPT and ChatDB, our DB-GPT fine-tuned severl commonly used LLMs for Text-to-SQL. By automating the query generation, DB-GPT enables the development of conversational agents
with advanced data analytics.

\section{Conclusion}
We presented an open-source, intelligent dialogue system for databases, which outperforms the best available solutions as evidenced by its superior capabilities in solving a wide range of tasks.
Our systematic approach contributes to the line of research on building LLMs for databases. In addition, our training and inference strategies may be useful for developing retrieval-based dialogue systems in general domains, allowing us to unlock broader real applications.

\bibliographystyle{icml2020_url}
\bibliography{references}  






\newpage
\appendix
\appendixpage

\section{Ongoing and Future Work}\label{app:future_work}

We are currently exploring several extensions to deal with more complex dialogue and analytics cases in our system. We are particularly interested in handling:
\begin{itemize}[leftmargin=*]

\item More powerful agents.  
Users may want our system not only to perform the analysis but also provide more powerful abilities, such as classical time series predictions~\citep{jin2023large,xue2021graphpp,xue2022hypro,xue2023easytpp} based on historical data and predictive decision abilities~\citep{xue_meta_2022,qu-2022-rltpp,pan2023deep}.

\item Integration of more model training techniques. In addition to pre-training, the community is also interested in continual learning techniques for language models, such as continual pre-training~\citep{jiang2023anytime}, prompt learning~\citep{wang2022learning,xue2023prompttpp}. The integration of these methods will greatly facilitate the research community in these areas.

\item More user-friendly presentation. 
Users may desire our system presenting answers in richer formats such as tables and diagrams. We have launched a new project DB-GPT-Vis\footnote{\small\url{https://github.com/eosphoros-ai/GPT-Vis}} that provides flexible and diverse visualization components for the chat box powered by LLMs.
\end{itemize}

\section{Experiment Details}

\subsection{Text-to-SQL Evaluation Details}
\label{app:text2sql_details}
\paragraph{Dataset Details.} \cref{tab:sql_dataset_details} shows the distribution of the dataset.

\begin{table*}[tbh]
	\begin{center}
		\begin{small}
			\begin{sc}
				\begin{tabularx}{1.00\textwidth}{l *{1}{S}*{5}{R}}
					\toprule
					Dataset & \multicolumn{5}{c}{\# Questions}  \\
					\cmidrule(lr){2-6}
					& Easy & Medium & Hard & Extra & Overall  \\
					\midrule		
 Spider-dev &  248   &               446   &               174       &           166              &    1034    \\   
 \bottomrule
				\end{tabularx}
			\end{sc}
		\end{small}
	\end{center}
	\caption{Text-to-SQL dataset details.}
	\label{tab:sql_dataset_details}
\end{table*}

\subsection{RAG Evaluation Details}
\label{app:rag_details}

\paragraph{Dataset Details.} We collect approximately $100$ questions approximately 100 questions each for the database domain (DatabaseQA) and the finance domain (FinancialQA). In addition, we annotate the questions by the difficulties proposed by experts. \cref{tab:rag_dataset_details} shows the statistics for both datasets.

\begin{table*}[tbh]
	\begin{center}
		\begin{small}
			\begin{sc}
				\begin{tabularx}{1.00\textwidth}{l *{1}{S}*{4}{R}}
					\toprule
					Dataset & \multicolumn{4}{c}{\# Questions}  \\
					\cmidrule(lr){2-5}
					& Easy & Medium & Hard & Overral  \\
					\midrule		
 DatabaseQA & 37	 & 35	& 16	& 88   \\   
 FinancialQA & 50	& 13	& 11	& 74   \\ 
 \bottomrule
				\end{tabularx}
			\end{sc}
		\end{small}
	\end{center}
	\caption{RAG dataset details.}
	\label{tab:rag_dataset_details}
\end{table*}

\subsection{SMMF Evaluation Details}
\paragraph{More Results.} We provide the results with Vicuna as the base LLM, shown in \cref{tab:main_results_smmf_vicuna}. The results are consistent with those in \cref{tab:main_results_smmf_qwen} and \cref{tab:main_results_smmf_baichuan}.

\begin{table*}[tbh]
	\begin{center}
		\begin{small}
			\begin{sc}
				\begin{tabularx}{1.00\textwidth}{l *{1}{S}*{1}{S}*{3}{R}}
					\toprule
					Model & \# Ccr & Platform & \multicolumn{3}{c}{Metrics}  \\
					\cmidrule(lr){4-6}
					& & & FTL (ms) & IL (s) & Throughput (tokens)  \\
					\midrule		
 Vicuna-7b & 4 & vLLM &  \textbf{23}&	\textbf{5}&	\textbf{217}   \\
 Vicuna-7b & 4 & HF &  815 & 67 & 15   \\
 \midrule
 Vicuna-7b & 16 & vLLM &  \textbf{36} &	\textbf{7}	& \textbf{646}   \\
 Vicuna-7b & 16 & HF &  5128 & 914 & 5   \\
 \midrule
 Vicuna-7b & 32 & vLLM &  \textbf{53}&	\textbf{8}&	\textbf{1000}   \\
 Vicuna-7b & 32 & HF &  11251 & 1453 & 6   \\      					\bottomrule
				\end{tabularx}
			\end{sc}
		\end{small}
	\end{center}
	\caption{Evaluation on SMMF with Vicuna as the base LLM.}
	\label{tab:main_results_smmf_vicuna}
\end{table*}

\section{Software Interface}\label{app:software}
\begin{figure*}[t]
    \includegraphics[width=0.99\linewidth]{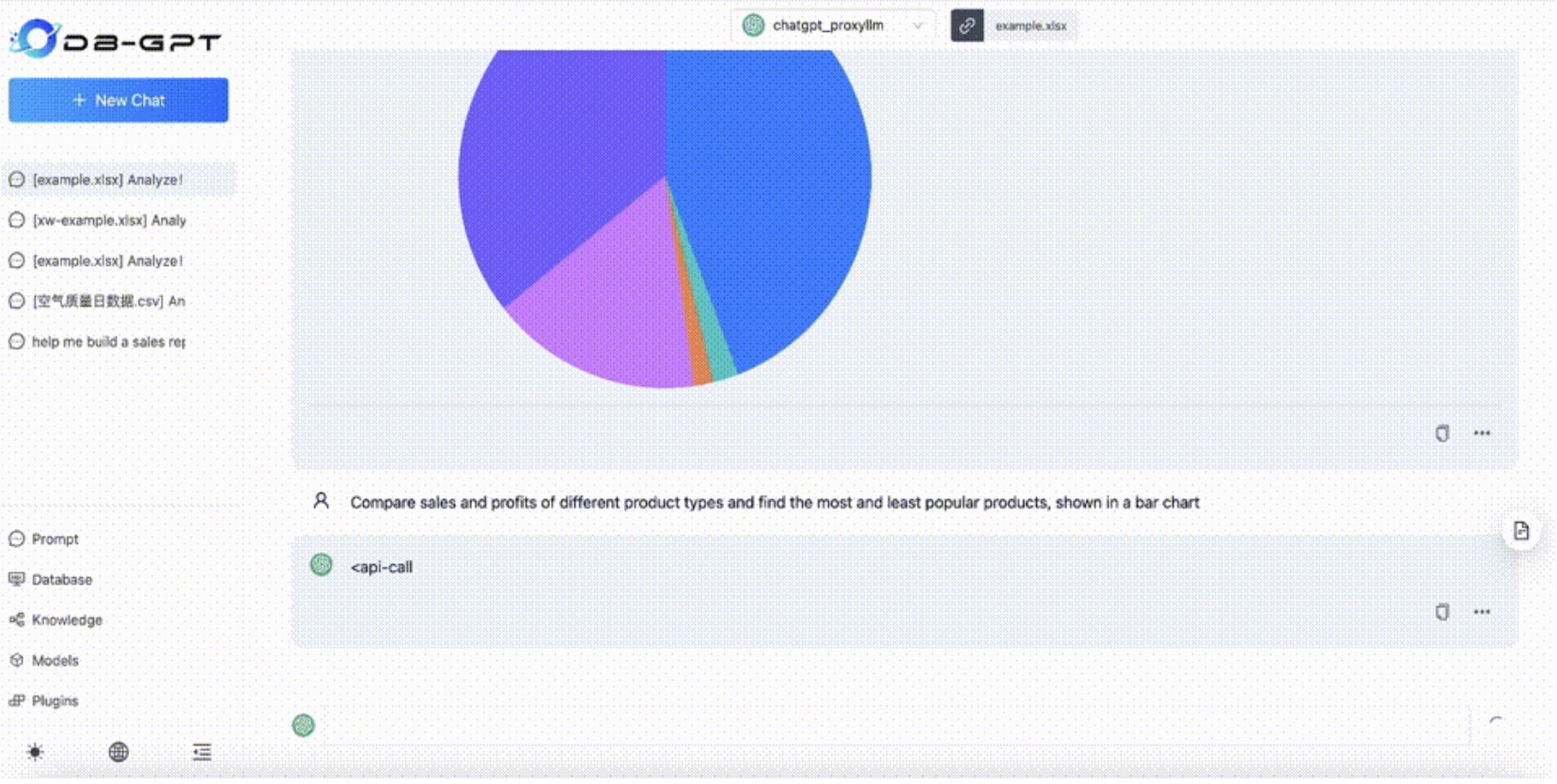}
    \caption{The main interface of DB-GPT: the configuration and chatbox.}
    \label{fig:webui}
\end{figure*}

\begin{figure*}[t]
    \includegraphics[width=0.99\linewidth]{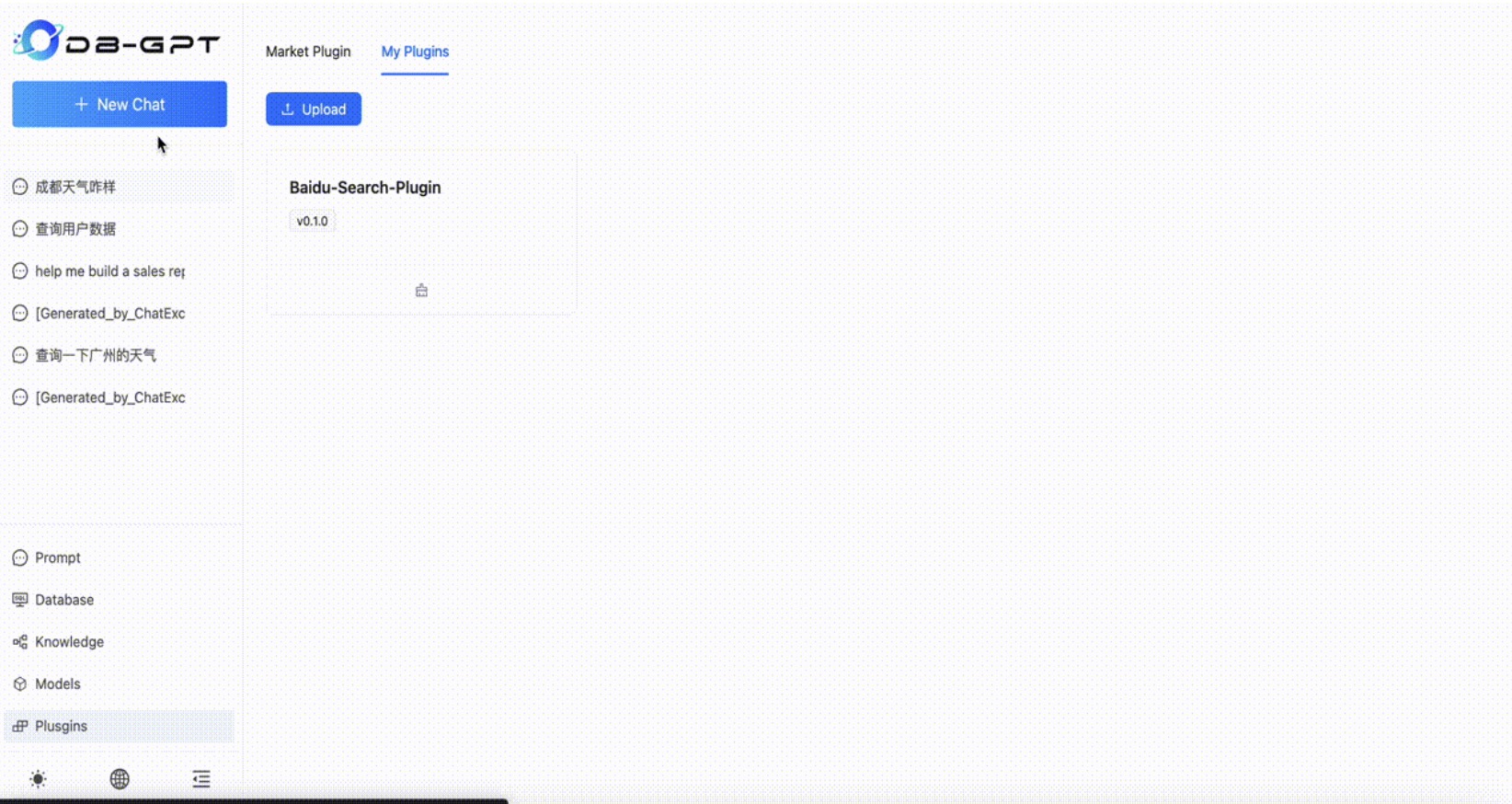}
    \caption{The 'plugin' tab of DB-GPT: the user can choose to load agent plugins (e.g., web search agent) for the QA task.}
    \label{fig:webui2}
\end{figure*}

The main interface of our DB-GP system can be seen in \cref{fig:webui,fig:webui2}. 

\end{document}